# Architecture of Network Management Tools for Heterogeneous System

Rosilah Hassan, Rozilawati Razali, Shima Mohseni,
Ola Mohamad and Zahian Ismail

Department of Computer Science,
Faculty of Information Science and Technology
Universiti Kebangsaan Malaysia, Bangi, Selangor, Malaysia
.

*Abstract*— **Managing heterogeneous network systems is a difficult task because each of these networks has its own curious management system. These networks usually are constructed on independent management protocols which are not compatible with each other. This results in the coexistence of many management systems with different managing functions and services across enterprises. Incompatibility of different management systems makes management of whole system a very complex and often complicated job. Ideally, it is necessary to implement centralized meta-level management across distributed heterogeneous systems and their underlying supporting network systems where the information flow and guidance is provided via a single console or single operating panels which integrates all the management functions in spite of their individual protocols and structures. This paper attempts to provide a novel network management tool architecture which supports heterogeneous managements across many different architectural platforms. Furthermore, an architectural approach to integrate heterogeneous network is proposed. This architecture takes into account both wireless fixed and mobile nodes.**

## I. INTRODUCTION

System and Network Management (S&NM) is concerned with observing, monitoring, testing, configuring, and troubleshooting the different network components, services, systems and users. The management process wraps all the network system elements starting from the end-users, through the applications and supporting data, the end system's network connectivity edge and deep into the network infrastructures themselves. Typically, S&NM comprises the following aspects:

- Human: where human manager defines the policy and organization approaches.
- Methodology: defines the architectural framework and the functions to be performed.
- Instrumentation: the actual operational aspects that establish the procedures, methods and algorithms for data collection, processing and reporting, and analysis of problems, their repair, prediction or forecasting of service levels and probable improvements to enhance performance.

S&NM aims to provide network managers a complete view of the whole network through a visualized Network Management Tool (NMT). The International Organization for Standards (ISO) [1] has categorized five main management functions that can be managed by such tools: Fault, Configuration, Accounting, Performance, and Security (FCAPS), as simplified in Table 1.

To illustrate how these functions are interrelated, an example of simple S&NM applications is shown in Fig. 1. It can be seen that a user interface is used to manage the functions, which are originated from various software, hardware, firmware, and end-users.

Most of the existing S&NM systems are developed in an individual fashion, where each system is designed to operate within its own defined area or scope.s This creates a number of incompatibilities and lack of integration does not allow a common view of the system and networks to be managed. Also, this causes lack of data flows between these incompatible systems, resulting in inconsistencies of data, event correlation and maintenance of the different data bases. It may also cause many systems with low level of inter and intra communications among them.

TABLE I.      ISO SPECIFIED FCAPS FUNCTIONS

| Category of Management | Task |
|---|---|
| Fault Management (FM) | Recognizing problem situation, isolating the problem to the source, Providing notification to the appropriate person, and tracking problems through resolution |
| Configuration Management (CM) | Controlling the health of network by keeping regular scheduled configuration back ups and having careful controlled implementation and also changing the procedures |
| Accounting Management (AM) | Measuring the usage statistics and allocation of costs associated with billing for time and services |
| Performance Management (PM) | Concerned with gathering network and system statistics including utilization, errors, response time are valuable tools in network trends |
| Security management (SM) | Controlling access rights, usually network infrastructure and access to network hardware components |

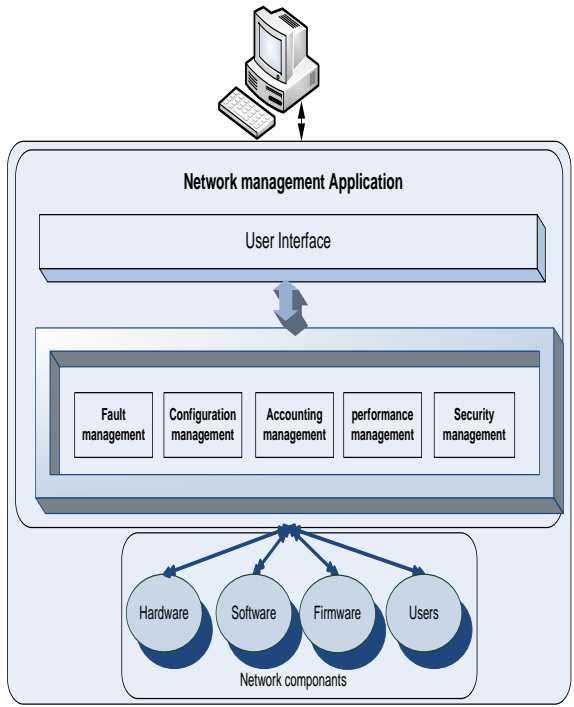

Figure 1.   Network management application components

Following are the most used protocols in network management systems:

- Simple Network Management Protocol.
- Telecommunication Management Network.
- Fault, Configuration, Accounting, Performance, and Security (FCAPS)

### A. Simple Network Management Protocol

Simple Network Management Protocol (SNMP) is an Internet Engineering Task Force (IETF) de facto standard known as Requests For Commands (RFC) [2]. SNMP is a framework for managing Internet devices or network elements using the TCP/IP protocol suite [3]. The SNMP management model contains a management station. The actual management process takes place in this management station. Other managed devices and available network peripherals communicate via a network management agent embedded inside the device with management station. The SNMP model of a managed network consists of four components: Managed nodes (Agent), Network Management stations (NMS), Managed information (MIB), and a management protocol (SNMP), as shown in Fig. 2.

The manager and managed agent style is used in SNMP where the manager monitors a group of agents. The dialogue relationship between manager and agent is shown in Fig. 3. A manager can check the agent behavior through a set of parameters; also it can force agent to behave in a certain way by resetting value of those parameters. Agents can send all or correlated alarms to the manager of any faulty situation. Different network managers can exchange information about each other's networks using SNMP. SNMP works in cooperation with a Structure of Management Information (SMI) and Management Information Base (MIB), as well as the underlying supporting S&NM protocol suite. SMI is responsible for defining the general rules for naming objects (hardware and system, non-physical such as programs, and administrative information), defining objects types, and show how to encode objects and values [4].

MIB is a conceptual database that can be distributed on different sites or assembled in a single location. It creates the objects needed and names them according to SMI rules. Then, it assigns them to their proper object types. For each entity such as a device, a set of defined objects will be created and kept in the database. Fig. 4 shows the ISO MIB that consists of the data which reflect the FCAPS that are requested by different network management architectures.

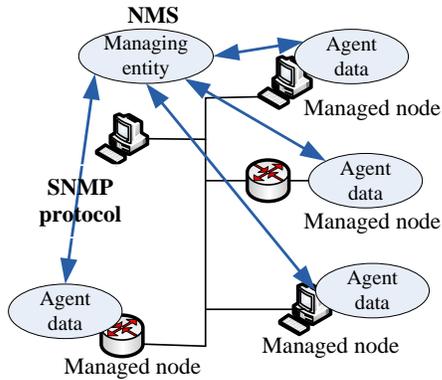

Figure 2.   SNMP management model components

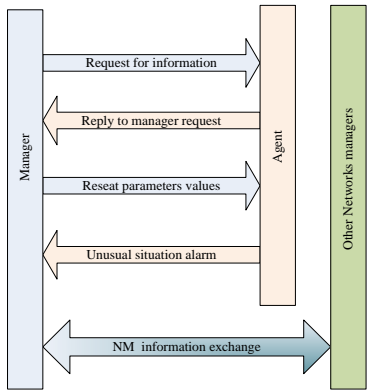

Figure 3.   Manager Agent Request Response Exchange Protocol

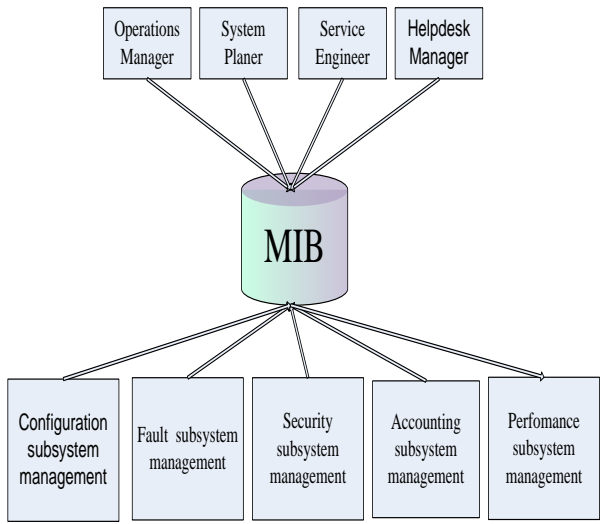

Figure 4.   ISO Management Information Base (MIB)

### B. Telecommunication Management Network

The Telecommunication Management Network (TMN) was defined by the International Telecommunication Union – Telecommunication Standardization Sector (ITU-T) [5]. It is a framework to achieve communication between heterogeneous telecommunication networks. TMN defines a standard interface for network elements that handle the communication process. In this way, network elements can be managed by a single network management system regardless of their different manufacturers. The framework identifies four logical layers of network management [5]:

- Network Management Layer (NML): Enables telecom operators to perform integrated fault management and service provisioning in multi-vendor multi-platform environments.

- Service Management Layer (SML): is where Telecom operators have sought to differentiate themselves by purchasing numerous applications for managing service usage, activation and assurance.

- Element Management Layer (EML): Contains individual network elements handling functions. Enables capabilities related to network monitoring, inventory management, service assurance, and network provisioning.

- Business Management Layer (BML): Performs functions related to business aspects, analyzes trends and quality issues provide a basis for billing and other financial reports.

Telecommunications technology architectures are typically expressed in a more simplified palette using the TMN model in Fig. 5.

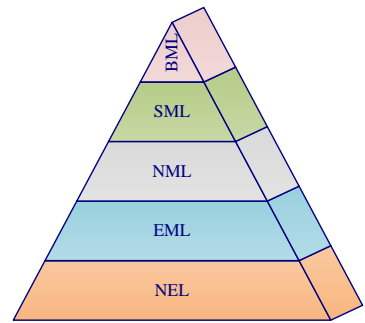

Figure 5.   TMN Network Management Architecture

## C. Fault, Configuration, Accounting, Performance, and Security (FCAPS)

FCAPS is a framework defined by the ISO and the name is a contraction of the five management categories that were mentioned earlier in Table 1: Fault, Configuration, Accounting, Performance, and Security.

This categorization is the functional one and does not consider the business-related role of management systems within the telecommunication network. The concept of FCAPS was initially developed by ITU-T to assist in managing the telecommunication networks. But, it was really the ISO who applied the concept to data networks. However, it turned out that those five protocols would be very similar, thus ISO decided to make them under one protocol, namely Common Management Information Protocol (CMIP) [6]. CMIP allows the communication between the manager and managed entities through a typical communication protocol. It can request the setting of parameter values, events and activities.

## II. NETWORK HETROGENITY

A heterogeneous network systems environment can be viewed as connecting computers and network devices such as switches, repeaters and routers with different protocols and different operating systems which varies in type, size, and topology. Fig. 6 illustrates three networks belonging to different autonomous organizations but providing a composite service to end-users.

The illustration in this figure can be described as being immature. Each network has a different environment and those environments can communicate with each other by a gateway. But in a real communication environment, the same device is surrounded by different technologies overlapping on the air such as WiFi technology cellular radio, Bluetooth, Zigbeee, and other various technologies with different capabilities and different coverage. The 3G devices, for example, support this variety and overlapping where user can have many types of connectivity using the same device in the same place at the same time. On the other hand, this is a mature heterogeneity.

The term heterogeneous in the near past means having different devices run by different operating systems to work under different technologies. This heterogeneity term is becoming less and less by embedding the variant capabilities in one device under one operating system, leading to more homogeneity.

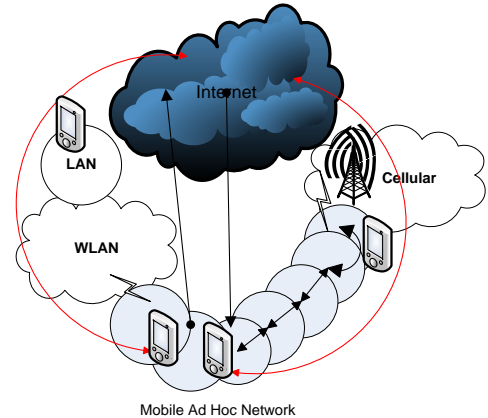

Figure 6. Example of Heterogeneous Network

It is the feature of modern communication world, where different service providers offering wide variety of services using different technologies, and targeting different user's interests. Forth generation (4G) technologies which bring high-speed broadband devices and services to the wireless world are good instances of heterogeneous networks where the user environment will be an integrated environment based on terminal heterogeneity and network heterogeneity. Heterogeneity could be derived from different issues as following:

- Terminal Heterogeneity: refers to the variety of user's terminals with different characteristics such as operating system, processing power, memory, storage size, battery life, technology supported.
- Network Heterogeneity: refers to the variety of network types and technologies used for example WiFi, Ethernet, Zigbee, Bluetooth, Cellular networks, GPS, etc.
- Protocol Heterogeneity: refers to the variety of technologies and network, where each type of network uses different communication protocols.
- Data type Heterogeneity: different networks, protocols, terminals can use different types of data.

## III. INTEGRATED NETWORK MANAGEMENT TOOL

In a heterogeneous environment, it becomes difficult for network managers to control all these networks or subnetworks for different reasons, such as incompatibilities or vendor specific S&NM architectures. There are also other reasons such as:

- Mastering problem: many management tools need to be mastered at least one for each type of device or network.

- Needs of expertise: an expert manager is needed to handle the heterogeneous system and network. Expert network managers are hard to find. Also, They are in short supply.
- Different protocols: subnetworks within the same network can use different suites of protocols.
- Data inconsistency: different subnetworks have different representation of data types and data flows.

Integrated S&NM needs to ensure that it creates a meta-level view and control of these different S&NMs. This need for a single centralized global view of a set of distributed components become essential in order to make management effective and efficient. Integrated System and Network management(IS&NM) was proposed to meet this need. The purpose IS&NM is to provide a single set of tools for managing network resources within a "networked set of components" regardless of the type of the subnets. IS&NM reduces errors in mastering multiple different subnetworks; minimizes the human skill levels needed to manage the network; and enhances network management flexibility simulatenously.

## IV. RELATED WORK

There are a number of network management tools that claimed to integrate network management functions. Those tools aimed to manage network components from the same vendor or the same service provider. Therefore, we are mainly interested in bulding a comprehensive network management tool through a single interface. We aim to provide a central control in a distributed management fashion. In this section, we will highlight several integrated network management architectures. Aslo, Table 2 summarizes three network management architectures and tools described is this section. Alcatel [7] offers a complete suite of network management systems for sub networks in Mobile Service Providers (MSP) environment. They can be integrated with Alcatel's mobile Operating Support System (OSS) solutions to provide full end-to-end service management. To help MSPs to meet these new challenges, Alcatel NMS and OSS have been designed as open solutions that facilitate the smooth integration of multi-technology and multi-vendor elements. Alcatel integrated fault management based on MicroMuse Netcool® consolidates real-time alarm information across network elements and application servers, providing real-time alerts relating to problems that affect service, thereby reducing the time to repair. Alcatel performance management based on Metrica™/NPR includes the collection of performance data and its consolidation across the entire network and application servers to provide end-to-end Quality of Service (QoS) indicators. Alcatel problem management based on Action Request System® from Remedy® integrates with previous applications, supporting network and service problem resolution by handling the workflow of problem resolution. It reduces network and service downtime and thus improves QoS and customer satisfaction. It can provide repair time and reaction time metrics as a basis for evaluating Service Level Agreements (SLA). The Evolium™ OMC-CN provides efficient cross-network element configuration capabilities and extensive performance monitoring functions tuned to the core network technologies. The Evolium™ OMC-R combines the integration of Radio Network Optimizer (RNO) with Alcatel's worldwide expertise in 2G mobile networks and its UMTS experience in Japan, to realize the best radio network management system on the market. In the CORDS project [8], several management tools have been developed independently. The tools are Network Modeling Tool (NetMod), Network Simulation Testbed (NEST), Network Management Analysis and Testing Environment (NETMATE), Hy+ and Shoshin Distributed Debugger.

TABLE II.    NETWORK MANAGEMENT ARCHITECTURE FOR ALCATEL, CORDS, AND NETDISCO

| Integrated Network Management & Tools | Network | In house Service Provider | Integrated Service Provider |
|---|---|---|---|
| Alcatel-Lucent | Mobile network | • Alcatel The Evolium™ OMC-CN<br>• Alcatel Evolium™ OMC-R | • MicroMuse Netcool®<br>• Metrica™/NPR<br>• Action Request System® from Remedy® |
| CORDS | Large-scale local network | Columbia University: NEST and NETMATE | University of Toronto: Hy+, University of Waterloo: Shoshin Distributed Debugger |
| Netdisco | Heterogeneous | - | Devices : Airespace, Allied Telesyn, Aruba, Asante, Bay, Cisco, Dell, Enterasys Networks, Extreme Networks, Foundry Networks, HP, Juniper, Net-SNMP, NetScreen, Nortel, Proxim, Sun, Synoptics and Zyxel.Software: Perl, Mason, Net-SNMP, PostgreSQL, Apache 1, GraphViz and MIBs |

NetMod is a software package that predicts the performance of new network technologies in a large-scale local network environment. NEST is a graphical environment for distributed network systems rapid prototyping and simulation developed at Columbia University. The NETMATE was developed at Columbia University to provide a unified and comprehensive software environment for network management to oversee and orchestrate the operations of diverse devices and protocol. Hy+ is a visual database system being developed at the University of Toronto. The system is capable of manipulating data by visually expressed queries on large complex systems such as a computer network. Shoshin Distributed Debugger is being developed at the University of Waterloo to support debugging of distributed and parallel applications

Network Discovery and Management (Netdisco) [9] is an Open Source web-based network management tool hosted by Sourceforge. Netdisco is a network management application targeted at large corporate and university networks. It integrated devices and software into one integrated system. The integrated devices are Airespace, Allied Telesyn, Aruba, Asante, Bay, Cisco, Dell, Enterasys Networks, Extreme Networks, Foundry Networks, HP, Juniper, Net-SNMP, NetScreen, Nortel, Proxim, Sun, Synoptics , and Zyxel [9]. These devices provide hardware such as switch, router and hub. The integrated softwares are Perl, Mason, Net-SNMP, PostgreSQL, Apache 1, GraphViz and MIBs [9].

## V. PROPOSED WORK

In this paper, we propose a network management tool architecture that supports heterogeneous network management system in many different architectural platforms. The proposed system should meet the following requirements.

- Easy to use: to minimize the level of expert needed to manage the network.
- Easy to access: web-based architecture is the best way to achieve this issue.
- The network management tool should have a standardized set of FCAPS function objects. This is needed to eliminate the data inconsistency that can be resulted from different standards in different networks.
- Able to interpret between definitions in different managed devices through a common nomenclature.

### A. Architectural Components

The proposed architecture is a collection of network management tools. The main components of this environment are the console, user interface, Agents Interface, and a database facility called MIB.

- User Interface (UI): This provides the ability to visually navigate and control complex network scenarios. The UI presents numerous objects and relationships into a visual representation for a human network manager. An important feature of user interface is that a user is able to see multiple views of network simultaneously. This feature of UI is shown in Fig. 7.
- Management Information Base (MIB): This part of architecture is containing information about objects (hardware and system, non-physical such as programs, and administrative information).MIB is organized by grouping of related objects and defines relationship between objects. Also, it is not a physical database; it is a virtual database that is complied into management model
- Agents Interface: the agents Interface of this architecture are entities providing information with a standardized interface for MIB. The gateway interacts with these agents interface asks the agents for information about the different network devices. These agents have a very simple structure and usually only communicate in response to the requests of variables contains in the MIB.
- Gateway: A gateway stands for communicating between networks that use different protocols, or which have the same protocols but do not otherwise communicate. Gateways integrate various data sources and create the appearance that all data resides in a single, logical relational database.

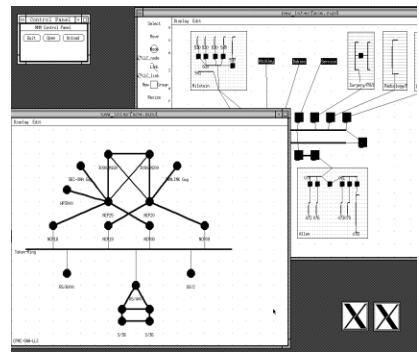

Figure 7.   Example of User Interface

### B. Process of proposed Integrated Network Management Tool

The process of proposed network management tool in this paper is as following. The local network management system in each network gathers the information needed about the devices, terminals, software and users in that network using its own protocols. The gathered data has to be sent to the Agents Interface in the Console part. The Agents Interface communicates with the agents and then stores those data in the MIB database. The Management Functions combines the FCAPS functions (Fault, Configuration, Accounting, Performance, and Security), which can be accessed by users through the User Interface. The user interface (UI) is a web-based interface that visualizes and shows the attached networks management information. In this architecture, agents on different networks interact with a gateway to communicate with any other agents. Networks such as LAN, GSM, ATM, and ISDN are local networks and each one has its own network management systems. Through gateway agent, each local network has an agent and they speak to gateway which leads to centralized control with distributed management. A conceptual view of the proposed architecture is shown in Fig. 8.

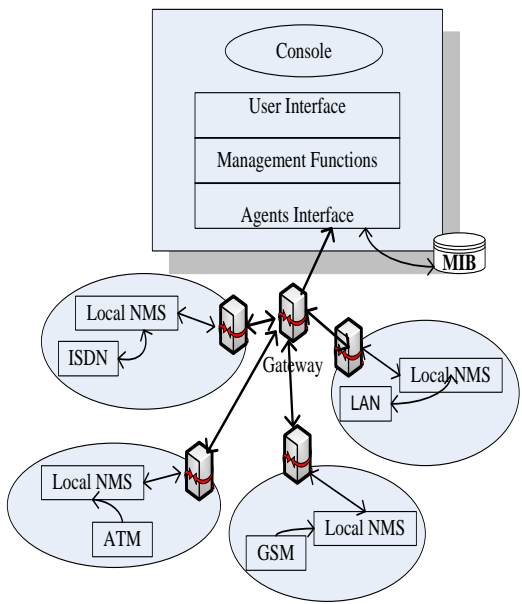

Figure 8.   The Conceptual View of the Proposed Network Management Tool  Architecture

### C.  Proposed Network Management Tool Requirements

There are four elements involved in the proposed management environment, as shown in Table 3.

The block diagram of a gateway application is shown in Fig. 9.  It consists of five components: Message Accepting Rely Agents, Message Extractor,

Message Translator, Message Rebuilder and Message Dispatching Relay Agents. The information received by the gateway from different networks would be processed as in Fig. 9.  In proposed architecture, the in-coming data from agents of different types of networks such as GSM, ATM, LAN, and ISDN are translated in the gateway agent. The information in the gateway agent is translated from actual to generic or vice versa. The functions of those elements are as follows:

- **Common Information Model (CIM):** This is an object oriented information model for specifying management information in a way that is independent of applications, platforms, protocols and implementations. CIM defines a way to exchange the data from any source and network. The data represented using CIM can be understood and analyzed by any network management tools or applications that understand CIM. The heterogeneous networks can make correlations between information coming from different locations in the network

- **Extensible Markup Language:** The Desktop Management Task Force (DMTF) recently advanced the CIM encoding to standard Extensible Markup Language (XML). In order to interoperate with each other for applications, it is necessary to represent actual management data in a standard way. Extensible Markup Language (XML) is a markup language for representing information in a standard format in order to make heterogeneous platforms and applications interoperate with each other.

- **CIM operation over HTTP:** This concept stands for mapping of CIM operations onto the Hyper Text Transfer Protocol (HTTP). All management functions can be accessed by users through the User Interface, which is a web based interface that visualize and show the attached networks management information.

- **Gateway:** A gateway which also described in the last section stands for communicating between two networks that use different protocols, or which have the same protocols but do not otherwise communicate. Gateways integrate various data sources and create the appearance that all data resides in a single, logical relational database. It is a combination of hardware and software that translates between two different protocols and acts as the connection point to the Internet.

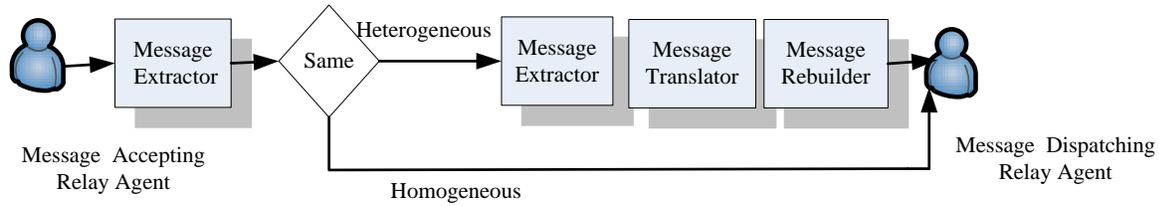

Figure 9.   Block Diagram of the Gateway Application

| Requirements | Application |
|---|---|
| Data description | Common Information Model (CIM) |
| lTransport Encoding | Gateway |
| Representing information | Extensible Markup Language |
| Operations to manipulate the data | Hyper Text Transfer Protocol (HTTP) |

## VI.   IMPLEMENTATION

We have developed an initial system prototype to proof of the proposed architecture and experimental development.  The prototype gives us a practical environment to evaluate and demonstrate our proposed framework in actual wireless IP networks. This prototype employs off-the-shelf fixed/mobile equipments and also, standardized interfaces to ensure interoperability with existing network and systems for pervasive applications.  The ad hoc network in this heterogeneous architecture incorporates on of the well-known MANET Reactive Protocols called AODV which is running on top of the IEEE 802.11 WLAN in ad hoc made to support multi-hop wireless communications. AODV-UU from Uppsala University is deployed as a user-space daemon that alters the kernel routing table dynamically as per the ever changing topology, and provides Internet connectivity through gateway support with tunnels [10].

### A. EXPERIMENTAL TESTBED AND EXPERIMENTS

We have developed an experimental work (Fig. 10) in our laboratory for performance evaluation of proposed architecture. Two main focuses in this experimental work are feasibility and usability. Without any infrastructure, using diverse terminal types (laptops, PDAs, and mobile phones) we were able to self-organize and form a community of group and/ private communications.

Various wireless platforms, namely Mobile Ad Hoc networks (MANET), Cellular networks (2G, 2.5G, 3G)[11], Wireless Local Area Networks WLANs (WiFi), Wireless Personal Area Networks (WPANs) like Bluetooth technology, zigbee, and ethernet networks like local area networks LAN are integrated in this experimental work. The widespread of the multi radio transmission mode wireless devices which are able to connect to a cellular base station CBS, and to a wireless access point WAP makes the daily used network an integration of different networks using different technologies. Much architecture has been proposed to integrate the wireless triangle [12]: Cellular, WLAN, and Mobile Ad Hoc networks. The proposed architecture consists of the following basic components (which are also summarized in Table 4):

**-Mobile nodes:** contains all the nodes that are free to move, which could be consist of the following:

- 3G Cellular phones that support WiFi, and GPS. The cellular phones will represent the dual mode nodes; Dual mode nodes have two applicability, they are MANETs mobile nodes and mobile gateway at the same time. Internet access to MANET nodes is provided through mobile gateways. The concept of mobile gateways is presented in [13].
- WLAN and Cellular network. Different battery life time is placed in each node.
- Laptops with WiFi and various battery life times.

**-Fixed nodes:** nodes are connected to fixed infrastructure, which is LAN (IP network) or to a cellular fixed infrastructure in case of CBS: like Pc connected to LAN

- Fixed Internet Gateway (IG):  IG works as an interface between two or more networks of different types. In this architecture, there will be one fixed gateway, and one of the cellular phones will act as the mobile gateway.  Both gateways have dual interfaces.
- Cellular Base Station (CBS): in this architecture assumption is that the coverage of the cellular base station is larger than the wireless access points used in WLANs at all times. This scenario uses only one CBS that covers area with 10km radios. This CBS is connected to a central CBS; this connection does not exist in Fig. 10.
- Three Wireless Access Point (WAP):  IEEE 802.11 access points with 100m radios coverage.

TABLE IV.    HETEREGENOUS NETWORK ARCHITECTURAL COMPONENTS

| | Number of nodes | Device | Single mode/ dual mode | Specifications |
|---|---|---|---|---|
| **Mobile Nodes** | 6 | cellular | Dual mode cellular and WLAN | 3G, movement speed 0m/S-20m/s , random directions |
| | 5 | laptop | Single mode WLAN | movement speed 0m/S-20m/s , random directions |
| **Fixed Nodes** | 1 | PC | LAN | Pentium III |
| | 1 | Internet Gateway IG | Dual mode | |
| | 1 | Cellular Base Station (CBS) | Single mode : cellular | 10km radios coverage |
| | 3 | Wireless Access Point (WAP) | Single mode : WLAN | 100m radios coverage, IEEE 802.11 |

—

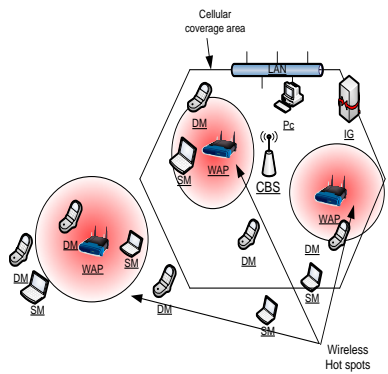

- CMS: cellular Mobile Station
- LAN: Local Area Network infrastructure
- WAP: Wireless Access Point
- DM: Dual Mode Node
- SM: Single Mode Node
- IG: internet Gateway

Figure 10.  Heterogeneous Network Testbed

## VII.  CONCLUSION AND FUTURE WORK

Diversity and complexity are the titles of the coming communication technologies. This situation caused by the increased production of the communication devices and systems without laying on one standardized concepts or common language. Most of the systems and devices nowadays concern on heterogeneous networks. That raised the intensive need to find one station to control and manage those networks, since controlling them separately brings a lot of difficulties and inconsistency. To conclude, we proposed network management tool architecture for managing heterogeneous networks with a web-based interface.  The architecture is being currently designed for implementation to meet the criteria as for IS&NM, which are easy to use, easy to access, and capable of interpreting different devices. To proof the proposed NM Tool we have developed a Testbed in our laboratory.  The result is an "Integrated Network Management Tool" targeting to enhance our future living environments.

The initial implementations demonstrate that proposed network management architecture provides a solid foundation for developing future network management tools. Our future work will be providing QoS to this architecture. We anticipate that this will enable supporting many services and applications in heterogeneous networks such as multimedia applications.

## ACKNOWLEDGMENT

This research is funded by Universiti Kebangsaan Malaysia (UKM) research projects UKM-GUP-NBT-08-29-116. The research group is known as Network Management Group. Please visit our website at http://www.ftsm.ukm.my/network for detail. Any opinions, findings and conclusions or recommendations expressed in this material are those of the authors.

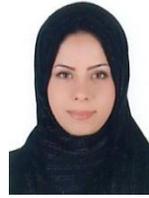

**Shima Mohseni** was born 1984 in Iran. She received her B.Sc. in computer science and Engineering from the Islamic Azad University of Najafabad, Iran in 2005. She is presently a M.Sc. student in the Information Technology department from faculty of Computer Science at Universiti Kebangsaan Malaysia. Also, she was working as research assistant with "network management tool group" in 2009. Her fields of interest include computer networking, mobile computing and QoS routing. She can be connected at shima.mohsenids@gmail.com

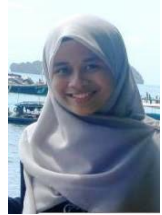

**Zahian Ismail** is a postgraduate student born in Malaysia. Her bachelor degree was in Computer Science and recently furthering her study in the same area at Faculty of Information Science and Technology in Universiti Kebangsaan Malaysia. She is a member of Network Management research group at her faculty. Her field of interest includes mobile ad hoc network, network communication, and QoS routing.

AUTHORS PROFILE

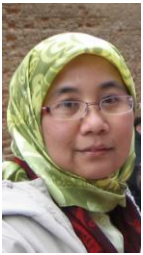

**Dr Rosilah Hassan** was born in Malaysia. She received her first degree from Hanyang University, Seoul, Republic of Korea in Electronic Engineering (1996). She work as an Engineer with Samsung Electronics Malaysia, Seremban before joining Universiti Kebangsaan Malaysia (UKM) in 1997. She obtains her Master of Electrical Engineering (M.E.E) in Computer and Communication from UKM in 1999. In late 2003 she went to Glasgow, Scotland for her PhD.   She received   her PhD in Mobile Communication from University of Strathclyde in 2008. Her research interest is in mobile communication, networking, 3G, and QoS. She is a senior lecturer at UKM for more than 10 years

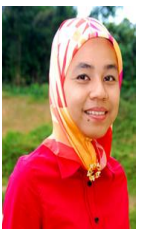

**Dr Rozilawati Razali** received her BSc. in Software Engineering from Sheffield Hallam University, United Kingdom in 1997. Prior to joining Universiti Kebangsaan Malaysia in 2003, she used to work in industry for about 6 years as a Software Engineer. She obtained her MSc. From Universiti Teknologi Mara in 2002 and PhD in Computer Science from University of Southampton, United Kingdom in 2008. Her research interests include software metrics, quality management and empirical software engineering.